\documentclass{article}
\usepackage[dvips]{epsfig}
\usepackage{graphicx}
\usepackage{latexsym,amsmath,amsfonts,amsthm,amssymb}

\usepackage[usenames]{color}

\usepackage[breaklinks,pdfpagemode=None,pdfview=FitH,pdfstartview=FitH,citebordercolor={0 0 1},linkbordercolor={0 0 1},urlbordercolor={0 0 1},pagebordercolor={0 0 1}]{hyperref}

\usepackage{graphics}

\theoremstyle{plain}
\newtheorem{theorem}{Theorem}

\newcommand{\fz}{{{z}}}

\newcommand{\rarrow}{\rightarrow}

\title{Dimensional Reduction and Crossover to Mean-Field Behavior for Branched Polymers}
\author{John Z. Imbrie\\
Department of Mathematics\\
University of Virginia\\
Charlottesville, VA  22904-4137\\}
\date{}
\begin{document}

\maketitle
\begin{abstract}
This article will review recent results on dimensional reduction for \linebreak branched polymers, and discuss implications for critical phenomena.  Parisi and Sourlas argued in \cite{PS81} that branched polymers fall into the universality class of the Yang-Lee edge in two fewer dimensions. Brydges and I have proven in \cite{BI01} that the generating function for self-avoiding branched polymers in $D+2$ continuum dimensions is proportional to the pressure of the hard-core continuum gas at negative activity in $D$ dimensions (which is in the Yang-Lee or $i \varphi^3$ class).   I will describe how this equivalence arises from an underlying supersymmetry of the branched polymer model.  

I will also use dimensional reduction to analyze the crossover of two-dimensional branched polymers to their mean-field limit, and to show that the scaling is given by an Airy function (the same as in \cite{C01}).
\end{abstract}

Let us begin with a definition of branched polymers, first on the lattice, then in continuous space.  On the lattice, branched polymers should look like the following picture.

\bigskip
\begin{center}
\begin{picture}(203,155)
\linethickness{2pt}

\put(-3.4,-4.75){
 \multiput(0,0)(0,30){6}{
  \multiput(0,15)(30,0){8}{${\cdot}$}
  }
}

\put(120,75){\line(1,0){60}}
\put(120,45){\line(0,1){60}}
\put(90,105){\line(1,0){60}}
\put(90,105){\line(0,1){30}}
\put(150,105){\line(0,1){30}}

\end{picture}
\end{center}

Define $c_N$ to be the number of $N$-vertex branched polymers mod translations. 
Let $T$ be a tree graph on $\{1,\dots,N\}$ and $y_1,\dots,y_N$ be an embedding of the graph in the lattice.  In other words, $y_i$ denotes the position of the $i^{\textrm{th}}$ vertex of the tree graph. If $i$ and $j$ are connected by a line in the tree graph, then $y_i$ and $y_j$ must be adjacent in the lattice.  In addition, no two vertices may be mapped to the same point; this would lead to a loop. The generating function $Z_{\rm{BP}}(\fz) = \sum_N \fz^N c_N$ can be written as
\begin{equation}
Z_{\rm{BP}}(\fz) = \sum^\infty_{N=1} \frac{\fz^N}{N!} \sum_T \sum_{y_{2},\ldots,y_{N}} \prod_{ij\in T} [2U'_{ij}]  \prod_{ij \notin T} U_{ij},
\end{equation}
where 
\begin{equation}
2U'_{ij} = \delta_{|y_{i}-y_{j}|,1}
\end{equation}
and 
\begin{equation}
U_{ij} = 1-\delta_{y_{i},y_{j}}
\end{equation}
enforce the adjacency and loop-free conditions, respectively.  For example, $c_3 = \frac{1}{3!} \cdot 3 \cdot 4 \cdot 3 = 6$ in $\mathbb{Z}^2$ as there are 3 tree graphs on \{1,2,3\}, and for each of them there are 4 choices for $y_2$ and 3 for $y_3$.  See \cite{BI01,F86} for details.
\subsection*{Background}
The concept of dimensional reduction was introduced by Parisi and Sourlas in 1979 in order to understand the behavior of the Ising model in a random magnetic field \cite{PS79}.  Although the proof of long-range order in three dimensions \cite{I84,I85} showed that dimensional reduction did not work in that problem, dimensional reduction appeared to work for branched polymers \cite{PS81}.  Our proof answers any lingering questions about dimensional reduction for branched polymers and gives a clear picture of how it works.  The question of what goes wrong for random fields continues to be debated in the literature \cite{BD98,PS02,F02}.

Dimensional reduction for branched polymers was employed recently by \linebreak Cardy to determine the crossover scaling function for area-weighted self-avoiding loops in two dimensions \cite{C01}.  Below, I will use our dimensional reduction identity to derive a similar result for the crossover from branched polymers to their noninteracting or mean-field limit.
\subsection*{Branched Polymers in $\mathbb{R}^{D+2}$}
Lattice models of branched polymers have been written down that are supersymmetric after dropping irrelevant terms \cite{S83,S85}.  Here we work in continuous space so that the supersymmetry responsible for dimensional reduction can be preserved.  Let us adjust the definitions given above for lattice branched polymers to the continuous case.  In $\mathbb{R}^{D+2}$, a branched polymer consists of

\begin{enumerate}
\item A tree graph $T$ on $1,\ldots,N$ and
\item An embedding $\mathbf{y}=\{y_i\}_{i=1,\ldots,N}$ into $\mathbb{R}^{D+2}$ such that
\begin{itemize}
\item if $ij\in T$ then $|y_{ij}|=1$ and
\item if $ij\notin T$ then $|y_{ij}|\ge 1$.
\end{itemize}
\end{enumerate}

The generating function for branched polymers (mod translations) is
\begin{equation}\label{eq4}
Z_{\rm{BP}}(\fz) = \sum^\infty_{N=1} \frac{\fz^N}{N!} \sum_T \int dy_2 \cdots dy_N \prod_{ij\in T} [2U'_{ij}] \prod_{ij \notin T} U_{ij},
\end{equation}
{where }
\begin{equation}
U_{ij}=U(|y_i-y_j|^2)
\end{equation}
{and }
\begin{equation}
2U'_{ij}=2U'(|y_i-y_j|^2)=\delta(|y_i-y_j|-1).
\end{equation}
As before, $U'_{ij}$ enforces adjacency, and $U_{ij}$ is a self-avoidance condition.  One can think of these branched polymers as collections of hard spheres with kissing conditions determined by the tree graph $T$.
\subsection*{The Hard-Core Gas in $\mathbb{R}^D$}
We shall relate branched polymers in $\mathbb{R}^{D+2}$ to the hard-core gas in $\mathbb{R}^D$.  The hard-core gas has hard spheres centered at $x_i \in \mathbb{R}^D$ for $i=1,\ldots,N$.  
Let
\begin{equation}
x_{ij}=x_i-x_j, \mbox{    } t_{ij}=|x_{ij}|^2,
\end{equation}
{and let}
\begin{equation}
U(t_{ij})=\theta(t_{ij}-1)
\end{equation}
enforce the hard-core constraint. Then
\begin{equation}
Z_{\rm{HC}}(\fz) = \sum^\infty_{N=0} \frac{\fz^N}{N!} \int_{\Lambda^N} dx_1\cdots dx_N 
\prod_{1 \leq i < j \leq N} U(|x_i-x_j|^2).
\end{equation}
\subsection*{Main Results}
\begin{theorem}\label{th1}
{For all $\fz$ such that the right-hand side converges absolutely,}
\begin{equation}
\lim_{\Lambda \nearrow \mathbb{R}^D} \frac{1}{|\Lambda|} \log Z_{\rm{HC}} (\fz) = -2 \pi Z_{\rm{BP}} \left( - \frac{\fz}{2\pi}\right).
\end{equation}
\end{theorem}
\subsubsection*{Examples}
Exact computation of the left-hand side is possible in $D=0,1$. For $D=0$, $Z_{\rm{HC}}(\fz)=1+z$ and so in dimension $D+2=2$,
\begin{equation}
Z_{\rm{BP}}(\fz)=-\frac{1}{2\pi} \log(1-2\pi \fz).
\end{equation}
The pressure of the one-dimensional gas of hard rods is also computable, see for example \cite{BI03}.  Hence in dimension $D+2=3$,
\begin{equation}
Z_{\rm{BP}}(\fz)=\frac{1}{2\pi} T(2\pi \fz) = \sum_{N=1}^{\infty} \frac{N^{N-1}}{2\pi N!}(2\pi\fz)^N.
\end{equation}
Here $T(z)=-\mathrm{LambertW}(-z)$ is the tree function \cite{CGHJK}.  These equations allow us to write a simple expression for $c_N$ in two and three dimensions.
\subsubsection*{Green's Functions}
We also have a dimensional reduction formula for correlations.  First, some definitions. 
Let
\begin{equation}\label{eq13}
\rho(\tilde{x}) = \sum^N_{i=1} \delta(\tilde{x} - x_i), \qquad 
\rho(\tilde{y}) = \sum^N_{i=1} \delta(\tilde{y} - y_i), 
\end{equation}
where $\tilde{x}, x_i \in \mathbb{R}^D$ and $\tilde{y}, y_i \in \mathbb{R}^{D+2}$. Then the density-density correlation functions of the two systems can be written as
\begin{eqnarray}
G_{\rm{BP}} (\tilde{y}_1,\tilde{y}_2;\fz)
& = & \sum^\infty_{N=1} \ \frac{\fz^N}{N!} \sum_T \int  dy_1 \cdots dy_N \, \rho(\tilde{y}_1) \rho(\tilde{y}_2) \prod_{ij \in T} [2U'_{ij}] \prod_{ij \notin T} U_{ij},
\nonumber \\[1mm]
G_{\rm{HC}} (\tilde{x}_1,\tilde{x}_2;\fz)
& = & \lim_{\Lambda \nearrow \mathbb{R}^{D}} \left\langle 
\rho(\tilde{x}_1) \rho(\tilde{x}_2)\right\rangle_{\rm{HC},\Lambda}.
\end{eqnarray}
Here $\langle \cdot \rangle_{\rm{HC},\Lambda}$ is the expectation in the measure for which $Z_{\rm{HC}}(\fz)$ is the normalizing constant.

\begin{theorem}\label{th2}
\begin{equation}
G_{\rm{HC}} (0,x;\fz)= -2 \pi \int d^2w G_{\rm{BP}} \left( 0,y;- \frac{\fz}{2\pi}\right),
\end{equation}
{where $y=(x,w) \in \mathbb{R}^{D+2}$.}
\end{theorem}

Analogous results can be obtained for the $n$-point functions.
\subsection*{Consequences for Critical Exponents}
One can define a critical exponent $\alpha_{\rm{HC}}$ by looking at the singularity of the pressure of the hard-core gas:
\begin{equation}
\lim_{\Lambda \nearrow \mathbb{R}^D} \frac{1}{|\Lambda|} \log Z_{\rm{HC}}(\fz)
\sim (\fz-\fz_c)^{2-\alpha_{\rm{HC}}}.
\end{equation}
Likewise, a susceptibility exponent $\gamma_{\rm{HC}}$ can be defined from the behavior of $Z_{\rm{BP}}$:
\begin{equation}
Z_{\rm{BP}}\left(-\frac{\fz}{2\pi}\right)
\sim (\fz-\fz_c)^{2-\gamma_{\rm{BP}}}.
\end{equation}
Keep in mind that $\fz_c$ is negative and $-\frac{z_c}{2\pi}$ is positive---Theorems \ref{th1} and \ref{th2} relate the critical behavior of branched polymers to the critical behavior of the hard-core gas at negative activity.  Theorem \ref{th1} equates the two singularities, so
\begin{equation}
\alpha_{\rm{HC}}=\gamma_{\rm{BP}}.
\end{equation}

Examining the 2-point functions of each model near their critical points, we can define exponents $\eta_{\rm{HC}}$, $\eta_{\rm{BP}}$ from the asymptotic forms
\begin{eqnarray}
G_{\rm{HC}}(0,x;{\fz}) &\sim& x^{-(D-2+\eta_{\text{HC}})}
K_{\text{HC}}(x/\xi_{\text{HC}}),
\\ [1mm]
G_{\rm{BP}}(0,x;{\fz}) &\sim& x^{-(d-2+\eta_{\text{BP}})}
K_{\text{BP}}(x/\xi_{\text{BP}}) 
\end{eqnarray}
as $|x|\rightarrow\infty$ with $\hat{x}:=x/\xi$ fixed.  Here $d=D+2$, $\xi_{\rm{HC}}$, $\xi_{\rm{BP}}$ are the correlation lengths of the respective systems, and $K_{\rm{HC}}$, $K_{\rm{BP}}$
are scaling functions.  Theorem \ref{th2} implies that $\xi_{\rm{HC}}=\xi_{\rm{BP}}$,
so in particular the exponents $\nu_{\rm{HC}}=\nu_{\rm{BP}}$, where
\begin{eqnarray}
\xi_{\rm{HC}}({\fz}) &\sim& (\fz-\fz_c)^{-\nu_{\rm{HC}}},
\\ [1mm]
\xi_{\rm{BP}}({\fz}) &\sim& (\fz-\fz_c)^{-\nu_{\rm{BP}}} 
\end{eqnarray}
as $\fz \searrow \fz_c$.
Theorem \ref{th2} also implies that 
\begin{equation}
{\eta_{\rm{HC}}=\eta_{\rm{BP}}}
\end{equation} 
for $D\ge 1$. Furthermore, if one
defines $\theta$ from
\begin{equation}
Z_{\rm{BP}}(\fz) = \sum_{N=1}^{\infty} c_N \fz^N
\end{equation}
{with}
\begin{equation}
c_N \sim \fz_c^{-N} N^{-\theta}.
\end{equation}
Then 
\begin{equation}
\theta = 3-\gamma_{\rm{BP}}.
\end{equation}

Further results can be obtained for the scaling functions.  We can express $K_{\rm{BP}}$
in terms of $K_{\rm{HC}}$ and its radial derivative:
\begin{equation}
K_{\text{BP}}(\hat{x}) = \frac{1}{4\pi^{2}}
\left[
\hat{x}K'_{\text{HC}}(\hat{x})-(D-2+\eta_{\text{HC}})K_{\text{HC}}(\hat{x})
\right].
\end{equation}
In $D=1$ we can compute \cite{BI03}
\begin{equation}
K_{\rm{HC}}(\hat{x})
=
-\frac{4}{\hat{x}^2} \ e^{-\hat{x}},
\end{equation}
so that in $D+2=3$ dimensions,
\begin{equation}
K_{\rm{BP}}(\hat{x})
=
\frac{1}{\pi^2 \hat{x}} \ e^{-\hat{x}},
\end{equation}
which agrees with the prediction of Miller \cite{M91}.
Let us tabulate the known or   for the critical exponents of the hard-sphere gas in various dimensions.  
\begin{center}
\begin{tabular}{|c|c|c|c|c|}
\hline
&&&&\\[-3.2mm]
 $D$ \mbox{\large{ }}  &$\hspace{4mm}\alpha\hspace{4mm}$& $\hspace{5mm}\nu_{\mathrm{HC}}\hspace{5mm} $& $\hspace{5mm}\eta_{\mathrm{HC}}\hspace{5mm}$ & $\sigma=1-\alpha$  \\[.8mm]
\hline
&&&&\\[-3.2mm]
 0  & ${2}$ & $\parbox[height=10mm]{1mm}{\huge{ } } $ &  & ${-1}$  \\[.8mm]
\hline
&&&&\\[-3.2mm]
 1 & ${\frac{3}{2}}$ & ${\frac{1}{2}}$ & ${-1}$ & ${-\frac{1}{2}}$  \\[.8mm]
\hline
&&&&\\[-3.2mm]
 2  & ${\frac{7}{6}}$ & $\frac{5}{12}$ & ${-\frac{4}{5}}$ & $-\frac{1}{6}$  \\[.8mm]
\hline
&&&&\\[-3.2mm]
 MFT $D>6$  & $\frac{1}{2}$ & $\frac{1}{4}$ & $0$ & $\frac{1}{2}$  \\[.8mm]
\hline
\end{tabular}
\end{center}
In $D=0,1$, the values are rigorous.  In $D=2$, the value $\alpha=\frac{7}{6}$ is obtained from Baxter's solution to the hard hexagon model \cite{B82,D83,BL87}.  The last column gives values for $\sigma$, the Yang-Lee edge exponent.  Assuming that the hard-sphere gas at negative activity falls into the Yang-Lee edge universality class, we have $\sigma=1-\alpha$.  Actually, from hyperscaling $(D\nu_{\rm{HC}}=2-\alpha_{\rm{HC}})$ and Fisher's relation $\sigma = \frac{D-2+\eta_{\rm{HC}}}{D+2-\eta_{\rm{HC}}}$, we know that there is only one independent exponent in each
dimension.  The value $\eta_{\rm{HC}}=-\frac{4}{5}$ is independently confirmed from conformal field theory \cite{C85}.

The branched polymer critical exponents in $D+2$ dimensions can be equated to the corresponding values for the hard sphere gas in $D$ dimensions; see the first three columns of the following table.
\begin{center}
\begin{tabular}{|c|c|c|c|c|}
\hline
&&&&\\[-3.2mm]
 $d=D+2$  & $\gamma=\alpha$ & $\nu_{\mathrm{BP}}=\nu_{\mathrm{HC}}$ & $\eta_{\mathrm{BP}}=\eta_{\mathrm{HC}}$ & $\theta=3-\gamma$  \\[.8mm]
\hline
&&&&\\[-3.2mm]
 2  & ${2}$ &  &  & ${1}$  \\[.8mm]
\hline
&&&&\\[-3.2mm]
 3  & ${\frac{3}{2}}$ & ${\frac{1}{2}}$ & ${-1}$ & ${\frac{3}{2}}$  \\[.8mm]
\hline
&&&&\\[-3.2mm]
 4  & $\frac{7}{6}$ & $\frac{5}{12}$ & $-\frac{4}{5}$ & $\frac{11}{6}$  \\[.8mm]
\hline
&&&&\\[-3.2mm]
 MFT $D>8$  & $\frac{1}{2}$ & $\frac{1}{4}$ & $0$ & $\frac{5}{2}$  \\[.8mm]
\hline
\end{tabular}
\end{center}
The values in two and three dimensions are now rigorous.  Note that in high dimensions ($d>8$) it has been proven that
$\gamma_{\mathrm{BP}}=\frac{1}{2}$, $\nu_{\mathrm{BP}}=\frac{1}{4}$,
$\eta_{\mathrm{BP}}=0$ (at least for spread-out lattice models)
\cite{HS90,HS92,HvS03}.  While our results do not apply
to lattice models, they give a strong indication that the
corresponding hard-core exponents have the same (mean-field) values
for $D>6$.
\subsection*{Forest-Root Formula}
The underlying mechanism behind all these results is an interpolation formula (the Forest-Root formula).  Let $f(\mathbf{t})$ depend on $\mathbf{t}=(t_{ij}), (t_i)$ for $1 \le i<j \le N$.
{Assume that $f\rightarrow 0$ at $\infty$.}
{Let $t_{ij} = |w_i-w_j|^2$, $t_i=|w_i|^2$ with $w_i \in \mathbb{C}$.}
Then
\begin{equation}
f (\mathbf{0}) = \sum _{(F,R)} \int_{\mathbb{C} ^{N}} f^{(F,R)}
(\mathbf{t})\left(\frac{d^{2}w}{-\pi } \right)^{N}.
\end{equation}
The sum is over {\textit{forests}} $F$ and {\textit{roots}} $R$ (collections of {\textit{bonds}} $ij$, and {\textit{vertices}} $i$, respectively) such that each tree of $F$ has exactly one root $R$.  See Fig 1.

\begin{figure}[ht]
\centering
\includegraphics[width=.97\textwidth]{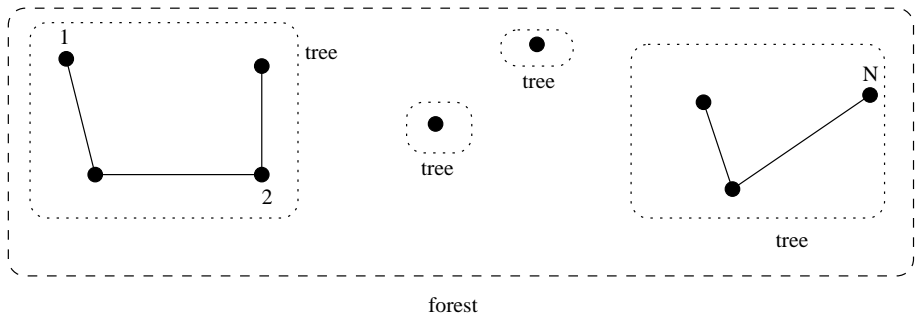}
\caption{Example of a forest}
\end{figure} 

For a simple example, consider the case  $N=1, F=\emptyset, R=\{1\}$.  Then the formula reduces to the fundamental theorem of calculus in one variable:
\begin{equation}
f(0)=\int_{\mathbb{C}} f'(t) \frac{d^2w}{-\pi} = -\int_0^{\infty} f'(t) dt.
\end{equation}
\subsection*{Supersymmetry}
To understand the Forest-Root formula in general, we need to exploit supersymmetry. 
Replace in $f(\bf{t})$ each variable $t_{i}$ with
\begin{equation}
 \tau_{i}=w_{i}\bar{w}_{i} + \frac{dw_{i} \wedge d\bar{w}_{i}}{2\pi i},
\end{equation}
and $t_{ij}$ with 
\begin{equation}
\tau_{ij}=w_{ij}\bar{w}_{ij} + \frac{dw_{ij} \wedge d\bar{w}_{ij}}{2\pi i}.
\end{equation}
{(Recall that $w_{ij}=w_i-w_j$).} Then
{$f(\underline{\tau})$ is defined by its Taylor series.}
I claim that a ``localization" formula holds:
\begin{equation}\label{eq34}
\int_{\mathbb{C}^N} f(\underline{\tau}) = f(\mathbf{0}).
\end{equation}
{This formula becomes the Forest-Root formula when expanded out.}
{The absence of loops comes from the fact that $(dw \wedge d\bar{w})^G = 0$ if $G$ has a loop.}
%The proof of the localization formula is nontrivial, and uses ideas from Witten's proof of the 
%Duistermaat-Heckmann theorem, hep-th/9204083.
We prove the formula by deforming the problem to the independent case, using ideas from \cite{W92}.

An alternate argument can be given, exploiting the linearity of (\ref{eq34}) to reduce to the case where $f$ is an exponential \cite{BW88}.  Let
\begin{equation}
\hat{f}(\mathbf{a})=\int_0^\infty d\mathbf{t} f(\mathbf{t})e^{\mathbf{a}\cdot\mathbf{t}}
\end{equation}
be the Laplace transform of $f$.  Here $\mathbf{a}=(a_i),(a_{ij})$ is dual to $\mathbf{t}$, that is $\mathbf{a}\cdot\mathbf{t} = \Sigma_{ij}a_{ij}t_{ij} + \Sigma_{i}a_{i}t_{i}$.  Let us assume that $f$ has exponential decay in each $t_{i}$ variable, so that we can take the $a_i$'s to have positive real part.  Cases where $f$ decreases more slowly at infinity can be treated by taking limits.  We can write $f$ in terms of its Laplace transform,
\begin{equation}
f(\mathbf{t}) = \int \prod_i \frac{da_i}{2\pi i} \prod_{ij} \frac{da_{ij}}{2\pi i} \hat{f}(\mathbf{a}) e^{-\mathbf{a}\cdot\mathbf{t}},
\end{equation}
with the integration contours running parallel to the imaginary axis.  We see that (\ref{eq34}) holds if it is true for functions of the form $e^{-\mathbf{a}\cdot\mathbf{t}}$.

Let us write
\begin{equation}
\mathbf{a}\cdot\mathbf{t} = 
\left<w,A\bar{w}\right> + \left<\frac{dw}{\sqrt{2\pi i}},A \frac{d\bar{w}}{\sqrt{2\pi i}}\right>,
\end{equation}
where
\begin{eqnarray}
A_{ij} &=& -a_{ij}, \qquad i\ne j,
\\[1mm]
A_{ii} &=& a_i + \sum_{j\ne i} a_{ij}.
\end{eqnarray}
The matrix $A$ is symmetric and has positive real part.  Then by the properties of the differential forms $dw_i$, $d\bar{w}_i$, 
\begin{eqnarray}
\int_{\mathbb{C}^N} e^{-\mathbf{a}\cdot\mathbf{t}} &=& 
\int_{\mathbb{C}^N} e^{-\left<w,A\bar{w}\right>} \exp \left[-\left<\frac{dw}{\sqrt{2\pi i}},A \frac{d\bar{w}}{\sqrt{2\pi i}}\right>\right]
\\[1mm]
&=&\int_{\mathbb{C}^N} e^{-\left<w,A\bar{w}\right>} \det{A} \prod_{i=1}^N \left[\frac{-dw_i \wedge d\bar{w}_i}{2\pi i} \right].
\end{eqnarray}
With $w_i = x_i + iy_i$, we have that $dw_i \wedge d\bar{w}_i = -2i dx_i \wedge dy_i$, so this is
\begin{equation}
\pi^{-N} \det{A} \int e^{-\left<w,A\bar{w}\right>} d^Nx\,d^Ny = 1,
\end{equation}
and we obtain (\ref{eq34}) for $f(\mathbf{t}) = e^{-\mathbf{a}\cdot\mathbf{t}}$.
\subsection*{Decoupling in Two Extra Dimensions}
Use the Forest-Root formula to decouple the spheres in the hard-core gas by moving them apart in two extra dimensions.
At fixed N we have an integral over $\mathbf{x} \in \Lambda^N \subset \mathbb{R}^{DN}$ of
\begin{equation}
f(\mathbf{0}) = \prod_{1\le i<j \le N} U(|x_{ij}|^2),
\end{equation}
where $t_i=|w_i|^2=0$.
Extend this to $w\ne 0$ by writing
\begin{equation}
f(\mathbf{t}) = \prod_{1\le i<j \le N} U(|x_{ij}|^2 + t_{ij}) \times (\mbox{large } t \mbox{ cutoff}).
\end{equation}
Apply the Forest-Root formula,
\begin{equation}
f (\mathbf{0}) = \sum _{(F,R)} \int_{\mathbb{C} ^{N}} f^{(F,R)}
(\mathbf{t})\left(\frac{d^{2}w}{-\pi } \right)^{N}.
\end{equation}
Then
each $\frac{d}{dt_{ij}}$, when applied to $U_{ij}$, becomes $\frac{1}{2}$ 
surface measure for the combined integrals over $y_{ij}=(x_{ij},w_{ij})$.  
This means that spheres are stuck together according to the forest $F$.
Furthermore, the trees of the forest decouple in the limit as the large $t$ cutoff is removed.
This is because they are spread apart in the $w$ directions, so that they rarely interact.  As a result, $Z_{\mathrm{HC}}$ can be written in the form of an exponential of tree graphs.  Thus
$\log Z_{\mathrm{HC}}$ is evaluated as a sum/integral over branched polymers, and Theorem \ref{th1} is proven.  See \cite{BI01} for details.
\subsection*{Relation with the Yang-Lee Edge}
Repulsive gases near criticality at negative activity are described by an $i\varphi^3$ field theory.  More precisely, they should fall into the same
universality class---see \cite{C82,D83,LF95,PF99}.
To see this, replace the hard-core potential with a repulsive, smooth, rapidly decaying two-body potential
\begin{equation}
u(x_{ij})=v(|x_{ij}|^2) \mbox{  with  } \hat{u}(k) > 0.
\end{equation}
By the sine-Gordon transformation, the repulsive gas
\begin{equation}
Z_v(\fz) = \sum_{N=0}^{\infty} \frac{\fz^N}{N!} \int_{\Lambda^N} dx_1 \cdots dx_N
\exp\left[-\beta \sum_{1 \le i<j \le N} u(x_{ij}) \right]
\end{equation}
can be written as a $-\tilde{\fz} e^{i\varphi}$ field theory.  With
$\tilde{\fz}=\fz e^{\beta v(0)/2}$, we have
\begin{equation}
Z_v(\fz) = \int \exp \left[\int_{\Lambda} dx \, \tilde{\fz} e^{i\varphi(x)} \right]
\frac{e^{-\tfrac12\left<\varphi,(\beta u)^{-1} \varphi \right>} [d\varphi]}{\cal{N}}.
\end{equation}

If $\tilde{\fz}$ is negative enough, this action looks critical.  In fact at the critical value of $\tilde{\fz}$, the lowest order term in the action is $i\varphi^3$.
A similar situation occurs for Ising or $\varphi^4$ model in a sufficiently large imaginary field \cite{F78}.
\subsection*{Dimensional Reduction: Soft Branched Polymers}
{The arguments leading to Theorems \ref{th1} and \ref{th2} work for any $U(t)$ for which $U-1$ and $U'$ have sufficient decay. } 
Take
\begin{equation}
U(t)=\exp(-\beta v(t))
\end{equation}
to get the identity
\begin{equation}\label{eq50}
\lim_{\Lambda \nearrow \mathbb{R}^D} \frac{1}{|\Lambda|} \log Z_v (\fz) = -2 \pi Z_{{\rm{BP}},v} \left( - \frac{\fz}{2\pi}\right),
\end{equation}
where $Z_{{\rm{BP}},v}$ is given by the same formula (\ref{eq4}) as before but with this $U$.
In this model of ``soft" branched polymers, the repulsion between monomers is given by a factor
{$\exp(-\beta v(|x_{ij}|^2))$}, and the attraction linking neighboring monomers in the tree graph is given by a factor
{$-2\beta v'(|x_{ij}|^2))$}.  
Together with the sine-Gordon representation for $Z_{{\rm{BP}},v}$, we obtain a direct representation for branched polymers in $D+2$ dimensions as a $-\tilde{\fz} e^{i\varphi}$ field theory in $D$ dimensions.
As this is presumably in the Yang-Lee class, we have confirmed the prediction of Parisi and Sourlas \cite{PS81}.
\subsection*{Crossover to Mean-Field Branched Polymers}
We wish to understand what happens as the repulsion between monomers is turned off.  In the limit, we obtain a mean-field or noninteracting model (on the lattice such models were analyzed in \cite{F86,BCvS}). Let $v=v(0)$ be small and $\fz$ be large so that a redefined
$\tilde{\fz}=\fz ve^{v/2}$ is fixed. In these variables (\ref{eq50}) becomes
\begin{equation}\label{eq51}
\log \int_{-\infty}^{\infty} \exp \left[-\frac{1}{v} \left(\tilde{\fz}e^{i\varphi}+
\tfrac12\varphi^2\right)\right]  \frac{d\varphi}{\sqrt{2\pi v}} = 
-2\pi Z_{{\rm{BP}},v}\left(\frac{\fz}{2\pi}\right).
\end{equation}
I have reversed the sign of $z$, so the branched polymer singularity is now at \textit{positive} $z$.  

It should be clear that the $v \rarrow 0$ limit simplifies both sides of this identity. On the 0-dimensional Yukawa gas side, $v \rarrow 0$ means the integral reduces to finding the critical point of $\tilde{z} e^{i\varphi} + \frac{1}{2} \varphi^2  $. On the two-dimensional branched polymer side, $v \rarrow 0$ eliminates the self-avoidance. 
Every tree has one fewer edges than vertices.  Assuming $v'$ remains proportional to $v$, as $v \rarrow 0$ and $zv \rarrow \tilde{z}$, this means that
$\frac{d}{dz}Z_{\mathrm{BP},v}$ goes to a finite limit,
which is a generating function of non-self-avoiding (or mean-field) branched polymers containing the origin.

We can observe the crossover by considering $Z_{\mathrm{BP},v}$ as a function of $\tilde{z}$ and $v$ near $v=0$ and $\tilde{z} = \tilde{z}_c$. With $S(\varphi) = \frac{1}{2} \varphi^2 + \tilde{z} e^{i\varphi}$, we have
\begin{equation}\label{eq52}
S'(\varphi) = \varphi + i \tilde{z} e^{i\varphi} = 0 \mbox{ when } i \varphi = x \mbox{ with } xe^{-x} = \tilde{z}
\end{equation}
\begin{center}
\begin{picture}(200,110)
\put(0,20){\line(1,0){200}}
\put(20,0){\line(0,1){100}}
\multiput(20,60)(9,0){9}{\line(1,0){5}}
\multiput(100,20)(0,8.8){5}{\line(0,1){5}}
\put(100,17){\line(0,1){5}}
\bezier{1000}(-10,-2)(40,35)(80,55)
\bezier{1000}(80,55)(100,65)(113,55)
\bezier{1000}(113,55)(140,25)(190,25)

\put(-11,58){$\scriptstyle{\tilde{z}_c = \, e^{-1}}$}
\put(90,11){$\scriptstyle{x_c = 1}$}
\put(7,103){$\scriptstyle{\tilde{z} = xe^{-x}}$}
\put(203,19){$\scriptstyle{x}$}

\end{picture}
\end{center}
The solution is $x=x(\tilde{z})=T(\tilde{z})$ as in the graph.  We also solve
\begin{equation}
S''(\varphi) = 1 - \tilde{z} e^{i\varphi} = 0\mbox{ when } \tilde{z} = e^{-i\varphi} = e^{-x} = \tfrac{\tilde{z}}{x},
\end{equation}
so $x_c = 1$, $\tilde{z}_c = e^{-1}$ has $S' = S''  = 0$.

Let us expand $S(\varphi)$ about $\varphi = -i$ with $\tilde{z}$ close to $\tilde{z}_c = e^{-1}$.  With $\tilde{\varphi}:=\varphi+i$ we obtain
\begin{eqnarray}
S(\tilde{\varphi}-i) & = & \tfrac12 \tilde{\varphi}^2 -i\tilde{\varphi} - \tfrac12 + e\tilde{z} \, e^{i\tilde{\varphi}} 
\nonumber\\[1mm]
&=& \tfrac12  \tilde{\varphi}^2 - i\tilde{\varphi} - \tfrac12 -e(z_c-z)e^{i\tilde{\varphi}} + e^{i\tilde{\varphi}}.
\end{eqnarray}
Let $t = \frac{\tilde{z}_c - \tilde{z}}{\tilde{z}_c}$. This becomes
\begin{equation}\label{eq55}
\left(e^{i\tilde{\varphi}} -1 - i\tilde{\varphi} + \tfrac12  \tilde{\varphi}^2\right)- t(e^{i\tilde{\varphi}}-1) - t + \tfrac12 
= -  \tfrac{ i \tilde{\varphi}^3}{6} -it \tilde{\varphi} + O(\tilde{\varphi}^4) + tO( \tilde{\varphi}^2) -t + \tfrac12,\end{equation}
and so from (\ref{eq51})
\begin{equation}
-2 \pi  Z_{\mathrm{BP},v} \left( \frac{z}{2\pi}\right) = \log  \int \exp \left[ \frac{1}{v} \left(\frac{i \tilde{\varphi}^3}{6} + it \tilde{\varphi} + O(\tilde{\varphi}^4) + tO(\tilde{\varphi}^2) + t - \frac12\right) \right] \frac{d \tilde{\varphi}}{\sqrt{2\pi v}}.
\end{equation}
Let us apply $\frac{d}{dz}$ to get the sum of branched polymers containing the origin. With $G_{\mathrm{BP},v}(y) = \frac{d}{dy} Z_{\mathrm{BP},v}(y)$, we have
\begin{equation}\label{eq57}
G_{\mathrm{BP},v} \left( \frac{z}{2\pi}\right) =- \frac{d}{dz} \log  \int \exp \left[ \frac{1}{v} \left(\frac{i \tilde{\varphi}^3}{6} + it \tilde{\varphi} + O(\tilde{\varphi}^4) + tO(\tilde{\varphi}^2) + t
\right) \right] d \tilde{\varphi}.
\end{equation}

The expression in the exponent is useful for $\tilde{\varphi}$ small.  Consider the crossover regime, $|t| \le O(v^{2/3})$.  I wish to show that the contribution to the integral from $|\tilde{\varphi}|>v^{1/3-\epsilon}$ is smaller than any power of $v$.  Make a kink in the contour of integration for $\tilde{\varphi}$, so that it lies at angles $\frac{\pi}{6}$ and $\frac{5\pi}{6}$. This turns $i \tilde{\varphi}^3$ into $-|\tilde{\varphi}|^3$.  
From (\ref{eq55}), one can check that on this contour, 
\begin{equation}
\mathrm{Re} \, S(\tilde{\varphi}) \ge \mathrm{const} \, \min\{|\tilde{\varphi}|^3,|\tilde{\varphi}|^2 \} 
-tO(\tilde{\varphi})-t + \tfrac12,
\end{equation}
so that $\tilde{\varphi}$ is effectively of order $v^{1/3}$.  This means that $\frac1v O(\tilde{\varphi}^4)$ and $\frac{t}{v}O(\tilde{\varphi}^2)$ are $O(v^{1/3})$ and that these terms do not contribute to 
$Z_{\mathrm{BP},v}$ to leading order in $v^{1/3}$.  Likewise when computing 
$\frac{d}{dz} = ve^{v/2} \frac{d}{d\tilde{z}} = -eve^{v/2} \frac{d}{dt}$ in (\ref{eq57}), the term
$O(\tilde{\varphi}^2)=O(v^{2/3})$ is subleading order.  The term $-\frac{d}{dz} \, \frac{t}{v}
=e^{1+v/2}$ is nonsingular.  Hence the singular behavior of $G_{\mathrm{BP},v}$ is
given to leading order by
\begin{equation}\label{eq59}
G_{\mathrm{BP},v}^{(\mathrm{sing})}\left(\frac{z}{2\pi}\right)=
- \,\frac{d}{dz} \log \int \exp\left[\frac{1}{v} \left( \frac{i \tilde{\varphi}^3}{6} + it \tilde{\varphi}\right)\right]d\tilde{\varphi}.
\end{equation}

We have arrived at a point very similar to the one Cardy reached in his analysis of the crossover from area-weighted self-avoiding loops to self-avoiding loops \cite{C01}.  He related that problem to branched polymers and, as we have done here, used dimensional reduction to compute the crossover scaling function as an Airy integral.  If we rescale $\varphi \rarrow \varphi v^{-1/3}$
in (\ref{eq59}), we obtain as in \cite{C01} that
\begin{equation}\label{eq60}
G_{\mathrm{BP},v}^{\rm (sing)} = v^{\frac{1}{3}} F(tv^{-\frac{2}{3}}),
\end{equation}
where
\begin{equation}\label{eq61}
F(s) = b_0 \frac{d}{ds} \ln \mbox{ Ai}(b_1s).
\end{equation}
Here $b_0$ and $b_1$ are constants, and Ai$(t) = \frac{1}{2\pi} \int^\infty_{-\infty} e^{it\varphi + i \varphi^3/3} d\varphi$ is the Airy function. 

The expression (\ref{eq60}) interpolates between two different critical behaviors for branched polymers.  We have already seen that $\gamma_{\mathrm{BP}}=1$ in two dimensions; this corresponds to the pole in (\ref{eq60}) that occurs when $v$ is fixed and nonzero, 
and $t$ decreases through $0$ to the first zero of Ai$(b_1t v^{-\frac{2}{3}})$.  
This is the limit of convergence of $Z_{\mathrm{BP}}$; past this point (\ref{eq51}) no longer makes sense.

If, instead, we send $v$ to zero and then let $t \searrow 0$, this means $\tilde{\varphi}$ is replaced by its critical value $\tilde{\varphi}_c$ where $S'(\tilde{\varphi}_c-i) = 0$, {\em c.f.} (\ref{eq52}). Then,
\begin{eqnarray}\label{eq62}
\lim_{v \rarrow 0} G_{\mathrm{BP},v} \left( \frac{z}{2\pi}\right)
& = & \lim_{v \rarrow 0} -ve^{v/2} \frac{d}{d\tilde{z}} \log  \int \exp \left[ -  \frac{1}{v} \left( \tilde{z} e^{i\tilde{\varphi}} + \tfrac12 \, \tilde{\varphi}^2\right)\right] d\tilde{\varphi}
\nonumber \\[1mm]
&  = &  e^{i\tilde{\varphi}_c}.
\end{eqnarray}
It is clear from the graph above that $i\tilde{\varphi}_c \sim -t^{1/2}$.  Thus the leading 
singularity of (\ref{eq62}) is $t^{1-\gamma}=t^{1/2}$, and $\gamma=\frac12$. This agrees with the value determined for the mean field models in \cite{F86,BCvS}. This also agrees with the conjectured value of the entropic exponent of self-avoiding walks. The square-root behavior can be seen in (\ref{eq60}) also, as $F(s) \sim s^{1/2}$ as $s \nearrow \infty$.

The identity of the crossover scaling functions and critical exponents for self-avoiding loops and branched polymers suggests a close connection between the two systems.  It is natural to think of loops arising as the frontier of branched polymers.  In this picture, the self-avoidance interaction of the branched polymer gives rise to an effective area weighting for the loop, and the mean-field limit of branched polymers corresponds to loops without area weighting.  This fits in with recent results \cite{LSW01} on the Hausdorff dimension of the frontier of Brownian motion (suggesting a link with self-avoiding walks).

\subsubsection*{Acknowledgement}

The author wishes to acknowledge conversations with David Brydges, John Cardy, Paul Fendley, and Yonathan Shapir, who helped shape my understanding of these problems.


\begin{thebibliography}{CGHJK}

\bibitem[Bax82]{B82}
R.~J. Baxter,
\newblock Exactly solved models in statistical mechanics,
\newblock Academic Press Inc. [Harcourt Brace Jovanovich Publishers], London,
  1982.

\bibitem[BCvS]{BCvS}
C.~Borgs, J.~Chayes, R.~van~der~Hofstad, and G.~Slade,
\newblock Mean-field lattice trees,
\newblock {\em Ann. Comb.} \textbf{3}, 205--221 (1999),
\newblock \href{http://arXiv.org/abs/math.PR/9904184}{arXiv:math.PR/9904184}.

\bibitem[BD98]{BD98}
E.~Br{\'e}zin and C.~De~Dominicus,
\newblock New phenomena in the random field {I}sing model,
\newblock {\em Europhys. Lett.} \textbf{44}, 13--19 (1998),
\newblock \href{http://arXiv.org/abs/cond-mat/9804266}{arXiv:cond-mat/9804266}.

\bibitem[BI01]{BI01}
D.C. Brydges and J.Z. Imbrie,
\newblock Branched polymers and dimensional reduction,
\newblock preprint, \href{http://arXiv.org/abs/math-ph/0107005}{arXiv:math-ph/0107005}.

\bibitem[BI03]{BI03}
D.C. Brydges and J.Z. Imbrie,
\newblock Dimensional reduction formulas for branched polymer correlation functions,
\newblock {\em J. Statist. Phys.} \textbf{110}, 503--518 (2003),
\newblock \href{http://arXiv.org/abs/math-ph/0203055}{arXiv:math-ph/0203055}. 

\bibitem[BL87]{BL87}
A.~Baram and M.~Luban,
\newblock Universality of the cluster integrals of repulsive systems,
\newblock {\em Phys. Rev. A} \textbf{36}, 760--765 (1987).

\bibitem[BW88]{BW88}
D.~C. Brydges and J.~Wright,
\newblock {M}ayer expansions and the {H}amilton-{J}acobi equation. {II}.
{F}ermions, dimensional reduction formulas,
\newblock {\em J. Statist. Phys.} \textbf{51}, 435--456 (1988).
\newblock Erratum: \emph{J. Statist. Phys.} \textbf{97}, 1027 (1999).

\bibitem[Car82]{C82}
J.~L. Cardy,
\newblock Directed lattice animals and the {L}ee-{Y}ang edge singularity,
\newblock {\em J. Phys. A} \textbf{15}, L593--L595 (1982).

\bibitem[Car85]{C85}
J.~L. Cardy,
\newblock Conformal invariance and the {Y}ang-{L}ee edge singularity in two
  dimensions,
\newblock {\em Phys. Rev. Lett.} \textbf{54}, 1354--1356 (1985).

\bibitem[Car01]{C01}
J.L. Cardy,
\newblock Exact scaling functions for self-avoiding loops and branched
  polymers,
\newblock {\em J. Phys. A} \textbf{34}, L665--L672 (2001),
\newblock \href{http://arXiv.org/abs/cond-mat/0107223}{arXiv:cond-mat/0107223}.

\bibitem[CGHJK]{CGHJK}
R.~M. Corless, G.~H. Gonnet, D.~E.~G. Hare, D.~J. Jeffrey, and D.~E. Knuth,
\newblock On the {L}ambert ${W}$ function,
\newblock {\em Adv. Comput. Math.} \textbf{5}, 329--359 (1996).

\bibitem[Dha83]{D83}
D.~Dhar,
\newblock Exact solution of a directed-site animals-enumeration problem,
\newblock {\em Phys. Rev. Lett.} \textbf{51}, 853--856 (1983).

\bibitem[Fis78]{F78}
M. E. Fisher,
\newblock {Y}ang-{L}ee edge singularity and $\varphi^3$ field theory,
\newblock {\em Phys. Rev. Lett.} \textbf{40}, 1610--1613 (1978).

\bibitem[Fel02]{F02}
D. E. Feldman,
\newblock Critical exponents of the random-field $O(N)$ model,
\newblock {\em Phys. Rev. Lett.} \textbf{88}, 177202 (2002),
\newblock \href{http://arXiv.org/abs/cond-mat/0010012}{arXiv:cond-mat/0010012}.

\bibitem[Fr\"o86]{F86}
J.~Fr{\"o}hlich,
\newblock Mathematical aspects of the physics of disordered systems.
\newblock In {\em Ph\'enom\`enes critiques, syst\`emes al\'eatoires, th\'eories
  de jauge, Part II (Les Houches, 1984)}, Amsterdam: North-Holland, 1986, pp. 725--893. 

\bibitem[HS90]{HS90}
T.~Hara and G.~Slade,
\newblock On the upper critical dimension of lattice trees and lattice animals,
\newblock {\em J. Statist. Phys.} \textbf{59}, 1469--1510 (1990).

\bibitem[HS92]{HS92}
T.~Hara and G.~Slade,
\newblock The number and size of branched polymers in high dimensions,
\newblock {\em J. Statist. Phys.} \textbf{67}, 1009--1038 (1992).

\bibitem[HvS03]{HvS03}
T.~Hara, R.~van~der Hofstad, and G.~Slade,
\newblock Critical two-point functions and the lace expansion for spread-out
  high-dimensional percolation and related models, 
\newblock {\em Ann. Probab.} \textbf{31}, 349--408 (2003),
\newblock \href{http://arXiv.org/abs/math-ph/0011046}{arXiv:math-ph/0011046}.

\bibitem[Imb84]{I84}
J.~Z. Imbrie,
\newblock Lower critical dimension of the random-field {I}sing model,
\newblock {\em Phys. Rev. Lett.} \textbf{53}, 1747--1750 (1984).

\bibitem[Imb85]{I85}
J.~Z. Imbrie,
\newblock The ground state of the three-dimensional random-field {I}sing model,
\newblock {\em Commun. Math. Phys.} \textbf{98}, 145--176 (1985).

\bibitem[LF95]{LF95}
S.~Lai and M.~E. Fisher,
\newblock The universal repulsive-core singularity and {Y}ang-{L}ee edge
  criticality,
\newblock {\em J. Chem. Phys.} \textbf{103}, 8144--8155 (1995).

\bibitem[LSW01]{LSW01}
G. F. Lawler, O. Schramm, and W. Werner,
\newblock The Dimension of the Planar Brownian Frontier is 4/3,
\newblock {\em Math. Res. Lett.} \textbf{8}, 401--411 (2001), 
\newblock \href{http://arXiv.org/abs/math.PR/0010165}{arXiv:math.PR/0010165}.

\bibitem[Mil91]{M91}
J.~D. Miller,
\newblock Exact pair correlation function of a randomly branched polymer,
\newblock {\em Europhys. Lett.} \textbf{16}, 623--628 (1991).

\bibitem[PF99]{PF99}
Y.~Park and M.~E. Fisher,
\newblock Identity of the universal repulsive-core singularity with
  {Y}ang-{L}ee edge criticality,
\newblock {\em Phys. Rev. E} \textbf{60}, 6323--6328 (1999),
\newblock \href{http://arXiv.org/abs/cond-mat/9907429}{arXiv:cond-mat/9907429}.

\bibitem[PS79]{PS79}
G.~Parisi and N.~Sourlas,
\newblock Random magnetic fields, supersymmetry and negative dimensions,
\newblock {\em Phys. Rev. Lett.} \textbf{43}, 744--745 (1979).

\bibitem[PS81]{PS81}
G.~Parisi and N.~Sourlas,
\newblock Critical behavior of branched polymers and the {L}ee-{Y}ang edge
  singularity,
\newblock {\em Phys. Rev. Lett.} \textbf{46}, 871--874 (1981).

\bibitem[PS02]{PS02}
G.~Parisi and N.~Sourlas,
\newblock Scale invariance in disordered systems:  the example of the random-field Ising model,
\newblock {\em Phys. Rev. Lett.} \textbf{89}, 257204 (2002),
\newblock \href{http://arXiv.org/abs/cond-mat/0207415}{arXiv:cond-mat/0207415}.

\bibitem[Sha83]{S83}
Y.~Shapir,
\newblock Supersymmetric dimer {H}amiltonian for lattice branched polymers,
\newblock {\em Phys. Rev. A} \textbf{28}, 1893--1895 (1983).

\bibitem[Sha85]{S85}
Y.~Shapir,
\newblock Supersymmetric statistical models on the lattice,
\newblock {\em Physica D} \textbf{15}, 129--137 (1985).

\bibitem[Wit92]{W92}
Edward Witten,
\newblock Two-dimensional gauge theories revisited,
\newblock {\em J. Geom. Phys.} \textbf{9}, 303--368 (1992),
\newblock \href{http://arXiv.org/abs/hep-th/9204083}{arXiv:hep-th/9204083}.

\end{thebibliography}
\end{document}